\documentclass[12pt]{article}

\usepackage{latexsym}
\usepackage{indentfirst}
\usepackage{amsmath}
\usepackage{amssymb}
\usepackage{graphicx}
\usepackage{bm}

\newcommand{\be}{\begin{equation}}
 \newcommand{\ee}{\end{equation}}
\newcommand{\bear}{\be\begin{array}}
\newcommand{\bea}{\begin{eqnarray}}
\newcommand{\eea}{\end{eqnarray}}

\begin{document}

\def\asq#1{\textbf{..(??. #1 .??)..} }          
\def\mem#1#2#3{  \left\langle #1 \left\vert  #2 \right\vert #3 \right\rangle   }

%
%
%
%
\title{Double lepton pair production with electron capture in relativistic heavy--ion collisions}


%
%
%
%

\author{
A. N. Artemyev\\ \small Helmholtz--Institut Jena, D--07743 Jena, Germany\\ \small St. Petersburg State Polytechnical University, St. Petersburg 195251, Russia
\and 
V. G. Serbo\\ \small Sobolev Institute of Mathematics, 630090 Novosibirsk, Russia\\ \small Novosibirsk State University, 630090 Novosibirsk, Russia
\and A. Surzhykov\\
\small Helmholtz--Institut Jena, D--07743 Jena, Germany
}



%
%
%
%




\maketitle

%
%
%
%

\begin{abstract}
We present a theoretical study of a double lepton pair production in ultra--relativistic collision between two bare ions. Special emphasis is placed to processes in which creation of (at least one) $e^+ e^-$ pair is accompanied by the capture of an electron into a bound ionic state. To evaluate the probability and cross section of these processes we employ two approaches based on (i) the first--order perturbation theory and multipole expansion of Dirac wavefunctions, and (ii) the equivalent photon approximation. With the help of such approaches, detailed calculations are made for the creation of two bound--free $e^+ e^-$ pairs as well as of bound--free $e^+ e^-$ and free--free $\mu^+ \mu^-$ pairs in collisions of bare lead ions Pb$^{82+}$. The results of the calculations indicate that observation of the double lepton processes may become feasible at the LHC facility.

%
%
%
\end{abstract}

%
%
%
%

\section{Introduction}
\label{Sec_Introduction}

Strong electromagnetic fields, induced in relativistic ion--ion as well as ion--atom collisions, may lead to a creation of electron--positron pairs. A large number of experimental and theoretical works have been performed to explore this quantum electrodynamic (QED) phenomenon \cite{BeG93,GuL99,BaH07}. However, while in the past most of the studies have dealt with single $e^+ e^-$ pairs, much of today interest is focused also on the \textit{multiple} pair production. Such a process, in which few electrons and positrons are created in a single collision event, is of high order in the electromagnetic coupling $\alpha$ and is predicted to have a sufficiently large cross section \cite{BaK02,LeM02}. Its analysis can help, therefore, to improve our understanding of the non--linear QED effects in quantum vacuum. Many attempts, based on coincidence observation of emitted electrons (or positrons), were made to detect the multiple production of \textit{free} $e^+ e^-$ pairs. Due to a large background signal these measurements were unsuccessful, thus indicating a need for an alternative scenario of the experiment. This scenario can be provided by the investigation of multiple $e^+ e^-$ creation events accompanied by the electron capture into bound ionic states. Even though the \textit{bound--free} pair production is much less probable than the free--free one, its experimental observation is feasible via detection of (one or few) down--charged ions. Such an experiment is likely to be performed at the LHC facility, whose very forward detectors allow efficient ion counting, and will reveal new information about the physics of extremely strong fields.

\medskip

In order to support future LHC experiments, we present a theoretical study of a \textit{double} lepton pair production in ultra--relativistic collisions between bare nuclei. Especially, we focus on the creation of two bound--free $e^+ e^-$ pairs in which electrons are captured into bound states of either the same:
\begin{eqnarray}
   Z_1 + Z_2 \, &\to& \, (Z_1+e^-+e^-)_{1s^2}+Z_2+e^+ +e^+ \, ,
   \label{eq:process_2ee_12} \\[0.2cm]
   Z_1+Z_2 \, &\to& \, Z_1+e^+ +e^+ + (Z_2+e^-+e^-)_{1s^2} \, ,
   \label{eq:process_2ee_21}
  \end{eqnarray}
or two different ions:
\begin{equation}
   Z_1 + Z_2 \, \to \, (Z_1+e^-)_{1s}+(Z_2+e^-)_{1s}+e^+ +e^+ \,.
   \label{eq:process_ee_ee}
\end{equation}
Here we assume, moreover, the most probable case when the ground ionic states are populated in course of the capture. The processes (\ref{eq:process_2ee_12})--(\ref{eq:process_2ee_21}) and (\ref{eq:process_ee_ee}) can be investigated experimentally by detecting residual helium--like or two hydrogen--like ions, and without the need for observation of positrons. These measurements can become feasible in the near future owing to (relatively) large corresponding cross sections which, being proportional to $\alpha^{14}$, may reach the order of 10~mb.

\medskip

Beside the reactions (\ref{eq:process_2ee_12})--(\ref{eq:process_ee_ee}), we consider---as the third scenario---the simultaneous production of a bound--free $e^+ e^-$ and a free--free $\mu^+ \mu^-$ pairs:
\begin{eqnarray}
   Z_1+Z_2 \, &\to& \, (Z_1+e^-)_{1s}+Z_2+\mu^++ \mu^-+e^+\, ,
   \label{eq:process_ee_mm_12} \\[0.2cm]
   Z_1+Z_2 \, &\to& \, Z_1+\mu^++ \mu^-+e^+ + (Z_2+e^-)_{1s} \,.
   \label{eq:process_ee_mm_21}
\end{eqnarray}
Even though the experimental study of this process is rather cumbersome and requires coincidence measurement of a down--charged ion and emitted muon (or anti--muon), it can be useful in the general discussion about electromagnetic processes at LHC. The theoretical background for the analysis of all three (double lepton) processes is discussed in Section~\ref{Sec_Theory}. We show, in particular, that the computation of the cross sections can be traced back to the probability of a pair production at a particular impact parameter. In the present work, two approaches were used to evaluate such a probability: apart from the first--order perturbation theory, we also employed analytical expressions derived from the equivalent photon approximation (EPA). Detailed calculations based on these two theories are presented in Section~\ref{Sec_Results_and_discussion} for collisions of bare lead nuclei Pb$^{82+}$ moving with Lorentz factors $\gamma$~=~1500 and 3000. For these typical LHC parameters, we predict up to ten thousands double pair production events per hour which can be measured by modern particle detectors. A brief summary of these results is given in Section \ref{Sec_Summary}.

\medskip

Relativistic units $c = \hbar = 1$ and $\alpha \approx 1/137$ are used throughout the paper unless stated otherwise.

%
%
%
%

\section{Theoretical background}
\label{Sec_Theory}

In order to study the (double) lepton pair production in ultra--relativistic heavy ion collisions we employ here the impact--parameter approach. Within such an approach, the projectile ion is assumed to move along a classical straight--line trajectory with an impact parameter $\rho$ and velocity $v$ as defined with respect to the target ion. The cross section of the pair creation is given then simply as:
\begin{equation}
   \label{eq:cross_section_general}
\sigma = \int\limits_{R_{1} + R_{2}}^{\infty} P(\rho) \, {\rm d}^2
\rho=2\pi \int\limits_{R_{1} + R_{2}}^{\infty} P(\rho) \,\rho\,
{\rm d} \rho \, ,
\end{equation}
where $R_1 + R_2$ is the sum of nuclear radii of colliding ions and the $P(\rho)$ is the probability of the process. As seen from this expression, the knowledge of the impact--parameter dependence of the $P(\rho)$ is crucial for the computation of the cross section $\sigma$. In the next sections we discuss, therefore, how the probability $P(\rho)$ can be evaluated for both the single and the double lepton pair production.

%
%

\subsection{Probability of a single $e^+ e^-$ pair production}
\label{Subsec_Single_pair_P}

\subsubsection{Equivalent photon approximation}

The creation of a \textit{single} electron--positron pair in ion--ion collisions has been intensively studied over the last decades (see Ref.~\cite{BaH07} and references therein). For ultra--relativistic energies, this process can be described by means of the equivalent photon approximation (EPA) which treats the electromagnetic field of a fast moving ion as a short pulse of linearly polarized light. The calculation of the $e^+ e^-$ cross section can be traced back, therefore, to the probability of pair \textit{photo}--production. For example, if collision of two nuclei with charges $Z_1$ and $Z_2$ is accompanied by a creation of a free positron and an electron in the ground state of the ``second'' ion
\begin{equation}
   \label{eq:single_production_general}
   Z_1 + Z_2 \to Z_1 + e^+ + (Z_2 + e^-)_{1s} \, ,
\end{equation}
the cross section of this process can be written as:
\begin{equation}
   \label{eq:single_production_cross_section}
d\sigma_{ee}(Z_2, Z_1) =  dn_{\gamma}(\omega_L,\rho) \, \,
\sigma_{\gamma}(\omega_L) \, .
 \end{equation}
Here, $dn_{\gamma}(\omega_L,\rho)$ is the number of (virtual) photons with energy $\omega_L$, produced by the ``first'' nucleus and \textit{seen} in the rest frame of the ``second'' nucleus. The $\sigma_{\gamma}(\omega_L)$ denotes then the cross section for the bound--free pair production following impact of the photon with energy $\omega_L$ on the ``second'' nucleus, $\omega_L + Z_2 \to e^+ + (Z_2 + e^-)_{1s}$.

\medskip

The evaluation of the cross section $\sigma_{\gamma}(\omega_L)$ and the photon number $dn_{\gamma}$ has been discussed previously in Ref.~\cite{LeM02,ArJ12}. In the present study, therefore, we will restrict ourselves to a short account of the basic formulas and ideas, which are needed to derive the $\sigma_{ee}$ and probability $P(\rho)$ of the bound--free pair production. Within the framework of the EPA and the leading logarithmic approximation, the number of equivalent photons for \textit{large} impact parameters, $1/m \ll \rho \ll \gamma_L/\omega_L$, reads as \cite{LeM02}:
\begin{equation}
   \label{eq:number_photons_2}
   dn_{\gamma}(\omega_L,\rho) = \frac{Z_1^2 \alpha}{\pi^2}\,
   \frac{d\omega_L}{\omega_L}\, \frac{d^2\rho}{\rho^2} \, ,
\end{equation}
where $\gamma_L$ is a Lorentz factor of a ``first'' nucleus in the rest frame of the ``second'' one. Inserting this expression into Eq.~(\ref{eq:single_production_cross_section}), we can find the cross section of the bound--free pair production:
\begin{eqnarray}
   \label{eq:single_production_cross_section_2}
   d\sigma_{ee}(Z_2, Z_1) &=& \frac{Z_1^2 \alpha}{\pi^2}\,
   \frac{d^2\rho}{\rho^2} \, \int_{2m}^{\infty}
   \frac{d\omega_L}{\omega_L} \,
   \sigma_{\gamma}(\omega_L) \nonumber \\[0.2cm]
   &=& \frac{Z_1^2 \alpha}{\pi^2}\,  \frac{d^2\rho}{\rho^2} \frac{274}{315} \, \frac{\alpha(Z_2\alpha)^5}{m^2} \, f(Z_2) \, ,
\end{eqnarray}
Here, the integration over the photon energy $\omega_L$ employs the \textit{explicit} form of the cross section $\sigma_{\gamma}(\omega_L)$ as obtained within the well--known Sauter approximation (see e.g. Eq.~(44) in Ref.~\cite{ArJ12}). The factor $f(Z_2)$, moreover, accounts for the difference between the Sauter predictions and rigorous relativistic calculations of the $e^+ e^-$ photo--production \cite{AgS97} and is inserted into
Eq.~(\ref{eq:single_production_cross_section_2}) to improve the accuracy of calculations. For the collisions of bare lead ions, which will be discussed in Section~\ref{Sec_Results_and_discussion}, this factor is $f(82) = 0.216$.

\medskip

By comparing the cross section (\ref{eq:single_production_cross_section_2}) with Eq.~(\ref{eq:cross_section_general}) one immediately derives the
probability of a single bound--free $e^+ e^-$ pair production:
\begin{eqnarray}
   \label{eq:single_production_probability}
   P_{ee}(\rho; Z_2, Z_1) = \frac{A(Z_2, Z_1)}{(m\rho)^2} \,,\;\;
\frac{1}{m} \ll \rho \ll \frac{\gamma_L}{\omega_L}\,,
\end{eqnarray}
where the coefficient $A$ depends only on the charges of colliding nuclei and is given by:
\begin{eqnarray}
   \label{eq:A_coefficient}
   A(Z_2, Z_1) = (Z_{1}\alpha)^2 (Z_{2}\alpha)^5 \,f(Z_{2}) \,
   \frac{274}{315\,\pi} \, .
\end{eqnarray}
These expressions describe the process (\ref{eq:single_production_general}) in which the electron is captured into the ground state of the ``second'' ion. Of course, the Eqs.~(\ref{eq:single_production_probability})--(\ref{eq:A_coefficient}) can be also applied to calculate the probability of the creation of an electron bound to the ``first'' nucleus; this would require just a substitution of charges $Z_1$ and $Z_2$.

\medskip

Equation~(\ref{eq:single_production_probability}) is valid for large values of $\rho \gg 1/m$, which give the main logarithmic contribution to the single pair--production cross section $d\sigma_{ee}$. In contrast, if few $e^+ e^-$ pairs are created in course of ion--ion collision, the region of \textit{small} impact parameters, $R_1+R_2< \rho\lesssim 1/m$, plays a significant role, as it will be shown in the next Section. Since
Eq.~(\ref{eq:single_production_probability}) is not justified in such a region, another approximation of $P_{ee}$ needs to be used. Based on the results of fully--relativistic perturbative calculations \cite{VaB84,BeG87}, which predict that the growth of the pair--production probability with decreasing $\rho$ slows down and eventually stops at a distance of a few Compton wavelengths, we conjecture that:
\begin{eqnarray}
   \label{eq:single_production_probability_small_rho}
   P_{ee}(\rho; Z_2, Z_1) \approx P_{ee}(1/m; Z_2, Z_1) = A(Z_2, Z_1) \, ,
\end{eqnarray}
for $R_1 + R_2 < \rho \lesssim 1/m$. The validity of such a na\"ive assumption will be discussed in Section~\ref{Sec_Results_and_discussion} where we show that calculations based on Eq.~(\ref{eq:single_production_probability_small_rho}) can reproduce the $e^+ e^-$ production cross sections with an accuracy of about 20 \%.

\subsubsection{First--order perturbation theory}
\label{Subsubsect_perturbative}

Beside the equivalent photon approximation, which will be used here for rough estimations, one can employ the first--order pertutbation theory and relativistic Dirac wavefunctions in order to calculate the probability $P_{ee}$. Within such an approach, the transition amplitude
\begin{eqnarray}
   \label{eq:transition_amplitude}
   a_{ee}(\rho) &=& i \gamma Z_1 e^2 \int {\rm d}t \, {\rm e}^{i \omega t} \, \int {\rm d}{\bm r} \, \psi_{e^-}^\dag({\bm r}) \,
   \frac{1 - v \alpha_3}{r'} \, \psi_{e^+}({\bm r}) \nonumber \\
\end{eqnarray}
is a ``building block'' from which all the properties of the bound--free pair production process (\ref{eq:single_production_general}) can be calculated \cite{VaB84,BeG87}. In this expression, $\omega = E_{e^+} + E_{e^-}$ is the sum of total energies of emitted positron and bound electron, and $\alpha_3 = \alpha_z$ is the Dirac matrix. Moreover, the $e^+ e^-$ pair production happens because of the Li\'enard--Wiechert potential $(1 - v \alpha_3)/r'$ of the ``first'' nucleus as seen in the rest frame of the ``second'' nucleus, and which depends on the time--dependent  distance $r' =
\sqrt{(x - \rho_x)^2 + (y - \rho_y)^2 + \gamma^2 (z - vt)^2}$.

\medskip

The computation of the transition amplitude (\ref{eq:transition_amplitude}) is significantly simplified if the standard radial--angular representation of the bound electron $\psi_{e^-}({\bm r})$ and positron $\psi_{e^+}({\bm r})$ functions is applied. For the continuum positron wave this representation can be achieved upon expansion of the $\psi_{e^+}({\bm r})$ into multipole (partial) components, characterized by a well--defined parity and total angular momentum. The partial--wave analysis is routine and has been performed in a large number of studies of various atomic processes \cite{VaB84,BeG87,SuF05,NaV09,EiS07}. In the present work, we used about 40 partial waves to calculate the amplitude $a_{ee}(\rho)$ and, then, the bound--free $e^+ e^-$ pair production probability:
\begin{eqnarray}
   \label{eq:single_production_probability_perturbative}
   P_{ee}(\rho; Z_2, Z_1) = \int {\rm d}{\bm p}_{+} \left| a_{ee}(\rho) \right|^2 \, ,
\end{eqnarray}
where the integration runs over the momentum ${\bm p}_{+}$ of emitted positron and the proper summation over the particle spins is implied (see Refs.~\cite{BeG87,SuF05} for further details).

%
%

\subsection{Probability of a double $e^+ e^-$ pair production}
\label{Subsec_Double_pair_P}

Based on the probability $P_e$, derived for the process (\ref{eq:single_production_general}), one can investigate also the \textit{double} bound--free pair production in a single collision between two nuclei. Two scenarios have to be considered for such a collision. In the first one, both electrons are captured by the same nucleus, c.f. Eqs.~(\ref{eq:process_2ee_12}) and (\ref{eq:process_2ee_21}). The probability of such a process can be expressed as a product of probabilities of a single pair production accompanied by the formation of (i) a hydrogen--like and, as a second step, (ii) a helium--like ion. If
the electrons are created in the ground state of the ion with a nuclear charge $Z_2$, this probability reads as:
\begin{equation}
   \label{eq:double_production_probability}
   P_{2ee}(\rho; Z_2, Z_1) = \frac{1}{2} \, P_{ee}(\rho; Z_2, Z_1) \,  P_{ee}(\rho; Z_2, Z_1) \, .
\end{equation}
Here, we neglected the interaction between bound electrons and just introduced the factor $1/2$ to account for the Pauli exclusion principle. Such an independent--particle model is usually well justified for atomic processes with heavy ions and has been successfully employed in a large number of studies \cite{SuJ06,SuJ08}.

\medskip

By inserting the probability $P_{2ee}$ into Eq.~(\ref{eq:cross_section_general}) one can derive the cross section
\begin{eqnarray}
   \label{eq:double_production_cross_section_general2}
   \sigma_{2ee}(Z_2,Z_1) &=&  \int_{R_1+R_2}^\infty P_{2ee}(\rho; Z_2, Z_1) \,  {\rm d}^2\rho \nonumber \\[0.2cm]
   &=& \frac 12 \int_{R_1+R_2}^\infty \left[ P_{ee}(\rho; Z_2, Z_1) \right]^2 \, 2\pi \rho \, {\rm
   d}\rho \,
\end{eqnarray}
of the process (\ref{eq:process_2ee_12}). While the evaluation of this cross section is the topic of Section~\ref{Sec_Results_and_discussion}, here we just note that the main contribution to the integral in Eq.~(\ref{eq:double_production_cross_section_general2}) is due to the region $\rho \lesssim 1/m$ since at $\rho \gg 1/m$ the integrand $\left[ P_{ee}(\rho; Z_2, Z_1) \right]^2 \, \rho \propto 1/\rho^3$ drops rapidly to zero. This observation confirms the importance of small impact parameters $\rho$ for the analysis of double pair production processes.

\medskip

In the second scenario, the ultra--relativistic collision between two nuclei leads to a formation of \textit{two} hydrogen--like ions in the process~(\ref{eq:process_ee_ee}), whose probability can be written as:
\begin{eqnarray}
   \label{eq:double_production_probability_2}
   P_{ee + ee}(\rho; Z_1, Z_2) &=& \frac{1}{2} \, \Big[ P_{ee}(\rho; Z_2, Z_1) \,  P_{ee}(\rho; Z_1, Z_2 - 1) \nonumber \\[0.2cm]
   && \hspace*{-1cm} + P_{ee}(\rho; Z_1, Z_2) \, P_{ee}(\rho; Z_2, Z_1 - 1) \Big]  \, .
\end{eqnarray}
Again, here we treat the double pair production as a two--step process. For instance, the first line of Eq.~(\ref{eq:double_production_probability}) represents the probability that (i) the virtual photon emitted by the nucleus $Z_1$ creates a bound electron and a positron in the field of bare ion with the charge $Z_2$, and (ii) the production of the second $e^+ e^-$ pair in the field of the ``first'' nucleus is induced by an impact with an electromagnetic field of the hydrogen--like ion $(Z_2 + e^-)_{1s}$. The factor $1/2$ accounts, moreover, for the identity of final states, as required by the Poisson statistics.

\medskip

In order to further simplify Eq.~(\ref{eq:double_production_probability_2}) we remind that small impact parameters, $\rho \lesssim 1/m$, provide the main contribution to the double--pair production cross cross sections. In this parameter range the nucleus of the hydrogen--like ion $(Z-1)_{1s}$ is weakly screened by the (bound) electron cloud with the mean square radius $\sqrt{\langle r^2\rangle} = \sqrt{3}/[(Z-1)\alpha m]> 1/m$. The electromagnetic field, produced by such an ion and as ``seen'' at $\rho \lesssim 1/m$ is very similar, therefore, to that of the bare nucleus $Z$. This allows us to approximate $P_{ee}(\rho; Z_{1,2}, Z_{2,1} - 1) \approx P_{ee}(\rho; Z_{1,2}, Z_{2,1})$, and to re--write
Eq.~(\ref{eq:double_production_probability_2}) as:
\begin{equation}
   \label{eq:double_production_probability_3}
   P_{ee+ee}(\rho; Z_1, Z_2) \approx  \, P_{ee}(\rho; Z_2, Z_1) \,  P_{ee}(\rho; Z_1,
   Z_2)  \,.
\end{equation}
In Section~\ref{Sec_Results_and_discussion} we will make use of this probability to investigate the cross section of the pair--production process (\ref{eq:process_ee_ee}).

%
%

\subsection{Probability of a free $\mu^+ \mu^-$ and bound--free $e^+ e^-$ pair production}
\label{Subsec_electron_muon_P}

Up to the present, we have applied the $P_{ee}$ to evaluate probabilities of various electron--positron production processes. The ultra--relativistic heavy ion collisions may lead also to the creation of \textit{other} leptons. Of special interest for the LHC physics, for example, are muon--antimuon pairs. While the production of single and double free--free $\mu^+ \mu^-$ pairs was considered in Ref.~\cite{HeK07}, here we study the process (\ref{eq:process_ee_mm_12})--(\ref{eq:process_ee_mm_21}) in which free--free $\mu^+ \mu^-$ and bound--free $e^+e^-$ pairs are created. If the electron is captured into a bound state of the ``second'' nucleus, the probability of such a process is
\begin{equation}
   \label{eq:electron_muon_production_P}
   P_{ee + \mu \mu}(\rho) = P_{\mu \mu}(\rho) \, P_{ee}(\rho; Z_2, Z_1) \, ,
\end{equation}
where $P_{\mu \mu}(\rho)$ is given, within the leading logarithmic approximation, by \cite{HeK07}:
\begin{equation}
   \label{eq:muon_production_probability}
   P_{\mu\mu}(\rho) = \frac{28}{9\pi^2}\,\frac{\left(Z_1 \alpha Z_2\alpha\right)^2}{(\mu\rho)^2}\, \Phi(\rho,\gamma)\,.
\end{equation}
In this expression, $\gamma$ is the Lorentz factor of the nuclei in the laboratory (collider) frame, $\mu$ is the muon mass, and the function $\Phi(\rho,\gamma)$ read as:
\begin{equation}
   \label{eq:Phi_function_1}
   \Phi(\rho,\gamma) = \left(4\ln{ \frac{\gamma}{\mu\rho}} +\ln{\frac{\rho}{R}}\right)\, \ln{\frac{\rho}{R}} \,
\end{equation}
for $R \ll \rho \le \gamma / \mu$, and
\begin{equation}
   \label{eq:Phi_function_2}
   \Phi(\rho,\gamma) =   \left(\ln{\frac{\gamma^2}{\mu^2  \rho R}}\right)^2
\end{equation}
for $\gamma / \mu \le \rho \ll \gamma^2 / (\mu^2 R)$.  In the past, approximate formulas (\ref{eq:muon_production_probability})--(\ref{eq:Phi_function_2}) were successfully employed to study muon--antimuon pair production in heavy--ion collisions. For the Pb--Pb scattering, for example, their predictions were found to be in a 10--15 \% agreement with the results of rigorous relativistic calculations \cite{HeT95}. Below, we will use the $P_{\mu\mu}(\rho)$ together with Eq.~(\ref{eq:single_production_probability}) and
(\ref{eq:electron_muon_production_P}) in order to calculate the cross section of the free $\mu^+ \mu^-$ and bound--free $e^+ e^-$ pair production.

\begin{figure}[t]
\vspace*{0.5cm}
\begin{center}
\includegraphics[width=12cm]{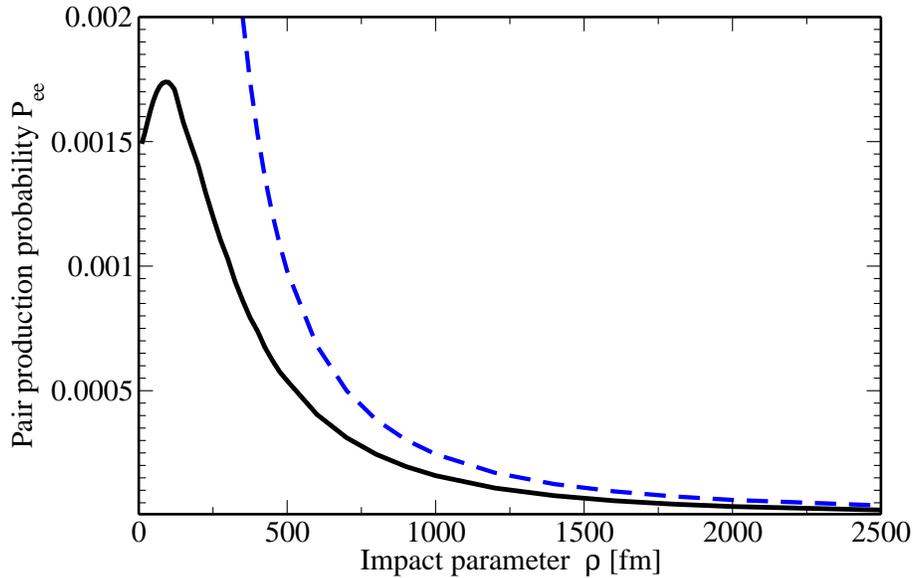}
\end{center}
\caption{\label{Fig1} The probability $P_{ee}$ of a single $e^+ e^-$ pair production in the collision of two led nuclei Pb$^{82+}$ accompanied by the electron capture into the ground ionic state. Results of the perturbative relativistic calculations (solid line) are compared with the predictions of Eq.~(\ref{eq:single_production_probability}), displayed by the dashed line. Calculations were performed for various impact parameters $\rho$ and ion energy $T_p$~=~100~GeV/u.
}
\end{figure}
%
%

%
%
%
%
\section{Results and discussion}
\label{Sec_Results_and_discussion}

As discussed in the previous section, the probabilities and, hence, the cross sections of the double lepton pair production processes~(\ref{eq:process_2ee_12})--(\ref{eq:process_ee_mm_21}) can be expressed in terms of the (single--pair) probability $P_{ee}$. Beside the standard first--order perturbative approach, we have also derived the analytical expressions from the equivalent photon approximation to find the $P_{ee}$. Before employing these EPA formulas for the estimation of the pair--creation cross sections, let us briefly discuss their validity both for small and large impact parameters $\rho$. In order to examine the performance of Eq.~(\ref{eq:single_production_probability}), used to describe the probability $P_{ee}$ for $\rho \gg 1/m$, we compared its predictions with those of the perturbation theory (\ref{eq:transition_amplitude})--(\ref{eq:single_production_probability_perturbative}). Perturbative calculations have been carried out for the collision of two lead nuclei Pb$^{82+}$ moving with the energy $T_p$~=~100~GeV/u in the laboratory frame. We restricted our analysis to this---relatively low---energy since the probability (\ref{eq:single_production_probability_perturbative}) is known to remain almost unchanged for $T_p \gtrsim 100$~MeV/u \cite{BeG87} and, moreover, a good convergence of the partial--wave expansion of the positron wavefunction can be achieved in such an (energy) region. As seen from Fig.~\ref{Fig1}, the perturbative results and approximation (\ref{eq:single_production_probability}) show better agreement as the impact parameter $\rho$ increases. For instance, while for $\rho = 386$~fm, which corresponds to the electron Compton wavelength $1/m$, Eq.~(\ref{eq:single_production_probability}) overestimates the results of rigorous relativistic calculations by more than factor of two, the discrepancy between the results is about 40~\% for $\rho = 2000$~fm. Such an accuracy is sufficient for the estimation of the pair--production cross sections as can be observed at the LHC facility.

\medskip

If for the large impact parameters, $\rho \gg 1/m$, the probability $P_{ee}$ of a single electron--positron pair production is approximated---within the EPA---by Eq.~(\ref{eq:single_production_probability}), in the region $2R < \rho \lesssim 1/m$ we use the na\"ive approach $P_{ee} \approx A$ with $A$ given by Eq.~(\ref{eq:A_coefficient}). For the ultra--relativistic collision between two lead nuclei the coefficient $A$ is as large as:
\begin{eqnarray}
   \label{eq:A_coefficient_Pb_Pb}
A\equiv   A(Z=82, Z=82) &=& \frac{274 \alpha^7}{315 \pi} (82)^7 \, f(Z = 82) \nonumber\\[0.2cm]
   &=& 1.65 \cdot 10^{-3} \,
\end{eqnarray}
which is in reasonable agreement with the results of the relativistic partial--wave theory obtained in the range $\rho \lesssim 400$~fm (cf. solid line in Fig.~\ref{Fig1}). Moreover, the approximation $P_{ee} \approx A$ can be also applied to estimate the \textit{cross section} of the $e^+ e^-$ pair production in the vicinity of nucleus. For instance, by performing integration in Eq.~(\ref{eq:cross_section_general}) from $\rho = 1/2m$ to $\rho = 1/m$ and we find
\begin{equation}
   \label{eq:cross_section_rho_interval}
   \sigma_{ee}(\Delta \rho) = \int_{1/2m}^{1/m} P_{ee}(\rho) {\rm d}^2\rho = \frac{3 \pi}{4 m^2} A = 5.8 \, \, {\rm barn}
\end{equation}
for the Pb$^{82+}$--Pb$^{82+}$ case. Comparison of this prediction with the result $\sigma_{ee}(\Delta \rho) \approx 4$~barn, obtained for the same range of impact parameters and within the non--perturbative coupled--channel theory \cite{BaR93}, again justifies the use of
Eq.~(\ref{eq:single_production_probability_small_rho}).

\medskip

Having proved the validity of Eqs.~(\ref{eq:single_production_probability}) and (\ref{eq:single_production_probability_small_rho}) we are ready
now to employ them---along with the perturbative theory---for the computation of the cross sections of double lepton pair production. Similar to before, we focus on the ultra--relativistic Pb$^{82+}$--Pb$^{82+}$ collisions and will study them for the scenario when two beams with Lorentz factors $\gamma_1=\gamma_2\equiv \gamma=1500$ and $3000$, as defined in the collider (center---of---mass) system, and the luminosity $\mathcal{L} = 10^{27}$~s$^{-1}$~cm$^{-2}$ move toward each other. As discussed already in Section~\ref{Subsec_Double_pair_P}, \textit{two} $e^+ e^-$ pairs can be created in these collisions accompanied by the capture of the electrons either (i) by the same (\ref{eq:process_2ee_12})--(\ref{eq:process_2ee_21}) or (ii) by two different (\ref{eq:process_ee_ee}) ions. By using Eqs.~(\ref{eq:cross_section_general}), (\ref{eq:double_production_probability}) and (\ref{eq:double_production_probability_3}), we easily find that the cross sections of these two processes are related to each other as:
\begin{equation}
   \label{eq:double_production_cross_section}
   \sigma_{ee+ee}=2\sigma_{2ee} = \int_{2R}^\infty
   \left[P_{ee}(\rho,Z,Z)\right]^2 \,2\pi \rho \, {\rm d}\rho
   \,,
\end{equation}
where $R_1 = R_2 = R$ is the radius of the lead nucleus. For the further evaluation of this expression, the knowledge about the $P_{ee}$ is needed. If, for example, the results of relativistic partial--wave calculations (cf. solid line in Fig.~\ref{Fig1}) are employed in Eq.~(\ref{eq:double_production_cross_section}), one finds, upon \textit{numerical} integration over the impact parameter:
\begin{equation}
   \label{eq:double_production_cross_section_result_1}
   \sigma_{ee+ee} = 2 \, \sigma_{2ee} = 11 \, \, {\rm mb} \, .
\end{equation}
This result can be compared also with the prediction based on the approximations (\ref{eq:single_production_probability}) and (\ref{eq:single_production_probability_small_rho}). Namely, since the $\left[ P_{ee}(\rho) \right]^2$ scales as $\sim 1/\rho^4$ for the large impact parameters (see Eq.~(\ref{eq:single_production_probability}) and Fig.~\ref{Fig1}), the main contribution to the integral in the right--hand side of (\ref{eq:double_production_cross_section}) originates from the region $\rho \lesssim 1/m$ where Eq.~(\ref{eq:single_production_probability_small_rho}) is conjectured to describe the $e^+ e^-$ production probability. If we employ this simple model in Eq.~(\ref{eq:double_production_cross_section}), we obtain:
\begin{equation}
   \label{eq:double_production_cross_section_2}
   \sigma_{ee+ee} = 2 \, \sigma_{2ee} \cong \int_{2R}^{1/m} A^2 \, \rho^2 \, {\rm d}\rho = \frac{\pi A^2}{m^2} = 12.6 \, \, {\rm mb} \, ,
\end{equation}
which is in reasonable agreement with the numerical relativistic result (\ref{eq:double_production_cross_section_result_1}). Our estimates of the cross sections $\sigma_{2ee}$ and $\sigma_{ee + ee}$ suggest that about 40000 double pair--production events (\ref{eq:process_ee_ee}) per hour may occur in course of Pb$^{82+}$--Pb$^{82+}$ ultra--relativistic collisions at the LHC. Moreover, one can observe approximately the same number of $e^+ e^-$ events in which the helium--like ions will be formed in either of colliding beams, i.e. when the process (\ref{eq:process_2ee_12}) \textit{or} (\ref{eq:process_2ee_21}) will take place.

\medskip

As mentioned already above, the bound--free $e^+ e^-$ pair might be also created together with the free muon and anti--muon, c.f. Eqs.~(\ref{eq:process_ee_mm_12}) and (\ref{eq:process_ee_mm_21}). The cross section of such a---rather exotic---process:
\begin{eqnarray}
   \label{eq:electron_muon_production_cross_section_1}
   \sigma_{ee + \mu\mu}(\gamma) &=& \int_{2R}^\infty P_{ee +
   \mu\mu}(\rho; Z, Z)
   \, \rho^2 \, {\rm d}\rho \nonumber \\[0.2cm]
   &=& \int_{2R}^\infty P_{\mu \mu}(\rho; Z, Z, \gamma)  \, P_{e e}(\rho; Z, Z) \,
    \rho^2 \, {\rm d}\rho \, ,
\end{eqnarray}
can be derived from Eqs.~(\ref{eq:cross_section_general}) and (\ref{eq:electron_muon_production_P}). In contrast to $\sigma_{2ee}$ and $\sigma_{ee+ee}$ this cross section exhibits an evident dependence on the Lorentz factor $\gamma$ which arises from the $\mu^+ \mu^-$ production probability (\ref{eq:muon_production_probability})--(\ref{eq:Phi_function_2}). In order to compute the $\sigma_{ee + \mu\mu}$, therefore, one has
to specify the energy of colliding ions. For example, by choosing the values $\gamma$~=~1500 and 3000, typical for LHC Pb$^{82+}$--Pb$^{82+}$ experiments, we find:
\begin{eqnarray}
   \label{eq:electron_muon_production_cross_section_2}
   \sigma_{ee + \mu\mu}(1500) = 2.2 \, \, {\rm mb} \,,\;\;
   \sigma_{ee + \mu\mu}(3000) = 2.6 \, \, {\rm mb} \, ,
\end{eqnarray}
which is comparable with the cross sections of the double free--free $\mu^+ \mu^-$ pair production, $\sigma_{2\mu \mu} \sim$~1~mb \cite{HeK07}. The results (\ref{eq:electron_muon_production_cross_section_2}) are based on the calculations of the $e^+ e^-$ production probability, performed within the framework of the perturbation theory
(\ref{eq:transition_amplitude})--(\ref{eq:single_production_probability_perturbative}). If, in contrast, the probability $P_{ee}$ is approximated by
Eqs.~(\ref{eq:single_production_probability}) and (\ref{eq:single_production_probability_small_rho}), one can derive for $Z_1 = Z_2 = Z= 82$:
\begin{eqnarray}
   \label{eq:electron_muon_production_cross_section_3}
\sigma_{ee + \mu\mu}(\gamma) &\sim& \int_{2R}^{1/m} A \, P_{\mu
\mu}(\rho; Z, Z, \gamma) \, 2 \pi\rho \, {\rm d}\rho
 \nonumber \\[0.2cm]
   &=& 6 \, A \, \frac{28}{27\pi} \, \frac{\left(\alpha Z \right)^4}{\mu^2} \,
    \int_{2R}^{1/m} \Phi(\rho, \gamma) \frac{{\rm d}\rho}{\rho}
   \nonumber \\[0.2cm]
   &=& 1.8 \, \, {\rm mb}\; (\mbox{for}\; \gamma =1500)
 \nonumber \\[0.2cm]
   &=& 2.1 \,  \, {\rm mb}\; (\mbox{for}\; \gamma =3000)  \, .
\end{eqnarray}
Again, both results are in a good agreement and imply that about $15\,000$ events (\ref{eq:process_ee_mm_12}) and (\ref{eq:process_ee_mm_21}) per hour can take place at the LHC collider. The detection of these events would require, however, a coincidence measurement the down--charged, hydrogen--like ions and emitted $\mu^-$ (or $\mu^+$). Due to the limited detector geometry, however, not all created muons can be registered and this might lead to a considerable reduction of the count rate.

\medskip

The above discussed calculations have been made under the assumption that an electron is captured into the ground ionic state of either hydrogen-- or helium--like heavy ions. The multiple $e^+ e^-$ pair production can be also accompanied by the formation of residual ions in their \textit{excited} states. Based on the previous studies \cite{ArJ12,LeM04}, we estimate that the contribution of such a excited--state recombination to the total cross sections $\sigma_{2ee}$, $\sigma_{ee+ee}$ and $\sigma_{ee+\mu\mu}$ does not exceed 25 \% which is well within the accuracy margin of the present calculations.

%
%
%
%
\section{Summary}
\label{Sec_Summary}

In summary, a theoretical study of double lepton pair production with an electron capture in ultra--relativistic heavy--ion collisions has been presented. Special emphasis was given to the processes involving creation of (one or two) bound--free $e^+ e^-$ pairs. In order to estimate the probabilities of these processes at different impact parameters we used two independent approaches. The first one is traced back to
the first--order perturbation theory and the partial--wave expansion of the Dirac wavefunctions, while the second employs simple analytical expressions derived within the framework of the equivalent photon approximation. Based on these approaches, calculations have been performed
for the typical LHC scenario in which two bare led ions Pb$^{82+}$, moving with the Lorentz factor $\gamma$~=~1500 and 3000, collide with each other. For such a collision, we analyzed the probabilities and, then, the cross sections of the double $e^+ e^-$ pair production accompanied by a formation of (i) a single helium--like (\ref{eq:process_2ee_12})--(\ref{eq:process_2ee_21}) or (ii) two hydrogen--like ions~(\ref{eq:process_ee_ee}), as well as (iii) the simultaneous creation of bound--free $e^+ e^-$ and free--free $\mu^+ \mu^-$ pairs~(\ref{eq:process_ee_mm_12})--(\ref{eq:process_ee_mm_21}). The predictions of the relativistic partial--wave theory and the EPA approach were found to be in a good agreement for all three processes, and have indicated that up to tens thousands of the (pair production) events per hour can happen in the course of high--$\gamma$ collisions at the LHC facility. With such a remarkable event rate the experimental studies of the above processes are likely to become feasible in the near future and will reveal new and unique information on the quantum electrodynamics in extremely strong electromagnetic fields.

%
%
%
%
\section*{Acknowledgements}

We are grateful to R.~Schicker who attracted our attention to this problem and explained us the details of the ALICE experiment. The stimulating discussions with A.~Milstein and A.~Voitkiv are also highly acknowledged. The work is supported by the ExtreMe Matter Institute (EMMI). A.~A. and A.~S. acknowledge support from the Helmholtz Gemeinschaft (Nachwuchsgruppe VH--NG--421). V.~G.~S. is supported by the Russian Foundation for Basic Research under the grant 13--02--00695.

%
%
%
%

\end{document}